\documentclass[pra,showpacs,showkeys,nofootinbib,twocolumn,10pt]{revtex4-1}
\usepackage{amsfonts,amsmath,amssymb,graphicx,epstopdf,verbatim,dsfont,color}
\usepackage[english]{babel}
\usepackage{subfigure,epstopdf}
\usepackage[normalem]{ulem}
\renewcommand{\vec}[1]{\mathbf{#1}}

\newcommand{\hk}[1]{{#1}}

\newcommand{\qq}{\textbf{q}}
\newcommand{\kk}{\textbf{k}}

\newcommand{\bb}[1]{\left( #1 \right)}
\newcommand{\bbcro}[1]{\left[ #1 \right]}

\renewcommand{\Re}{\textrm{Re}}
\renewcommand{\Im}{\textrm{Im}}
\begin{document}

\title{Prethermalization to thermalization crossover in a dilute Bose gas following an interaction ramp}
\author{Mathias Van Regemortel}
\email{Mathias.VanRegemortel@uantwerpen.be}
\affiliation{TQC, Universiteit Antwerpen, Universiteitsplein 1, B-2610 Antwerpen, Belgium}
\author{Hadrien Kurkjian}
\affiliation{TQC, Universiteit Antwerpen, Universiteitsplein 1, B-2610 Antwerpen, Belgium}
\author{Iacopo Carusotto}
\affiliation{INO-CNR BEC Center and Dipartimento di Fisica, Universit\`a di Trento, via Sommarive 14, 38123 Povo, Italy}
\author{Michiel Wouters}
\affiliation{TQC, Universiteit Antwerpen, Universiteitsplein 1, B-2610 Antwerpen, Belgium}
\date{\today}

\begin{abstract}
The dynamics of a weakly interacting Bose gas at low temperatures is close to integrable due to the approximate quadratic nature of the many-body Hamiltonian. While the short-time physics after an abrupt ramp of the interaction constant is dominated by the integrable dynamics, integrability is broken at longer times by higher-order interaction terms in the Bogoliubov Hamiltonian, in particular by Beliaev-Landau scatterings involving three quasiparticles. The two-stage relaxation process is highlighted in the evolution of local observables such as the density-density correlation function: an \hk{integrable} dephasing mechanism leads the system to a prethermal stage, followed by true thermalization conveyed by quasiparticle collisions. Our results bring the crossover from prethermalization to thermalization within reach of current experiments with ultracold atomic gases.

\end{abstract}
\maketitle

\section{introduction}
Ever since the development of quantum mechanics in the early days, it has been a central question how the unitary time evolution of a quantum wavefunction of many particles may generate a seemingly thermal ensemble in the long-time limit -- at least in the eyes of an experimenter with limited tools to probe the system. The \emph{eigenstate thermalization hypothesis} (ETH) \cite{deutsch_ETH,srednicki_ETH} aims to address this question by stating that expectation values of macroscopic observables computed with respect to a single generic eigenstate of energy $E$ are the same as the microcanonical average around the corresponding energy. The hypothesis has been verified numerically for a wide series of chaotic quantum systems \cite{rigol_ETH,rigol_ETH2}.

Since it relies on the hypothesis of ergodicity, ETH is not expected to hold for integrable quantum systems. There, an extensive number of conserved quantities restricts the full quantum dynamics to a small subspace of the total \hk{phase} space, thereby preventing thermalization. The long-time states of integrable systems can still be statistically described by a stationary \emph{generalized Gibbs ensemble} (GGE) \cite{rigol_GGE}, that incorporates all the conserved charges, as recently seen in a cold atom experiment \cite{schmiedmayer_GGE}. Another seminal experimental example is the quantum Newton cradle \cite{kinoshita_newton}. In the same spirit as ETH, a representative eigenstate of the integrable Hamiltonian can be identified based on these conserved charges, which correctly reproduces expectation values of local observables \cite{caux_quench}.

\begin{figure}
\centering
\includegraphics[width=\columnwidth]{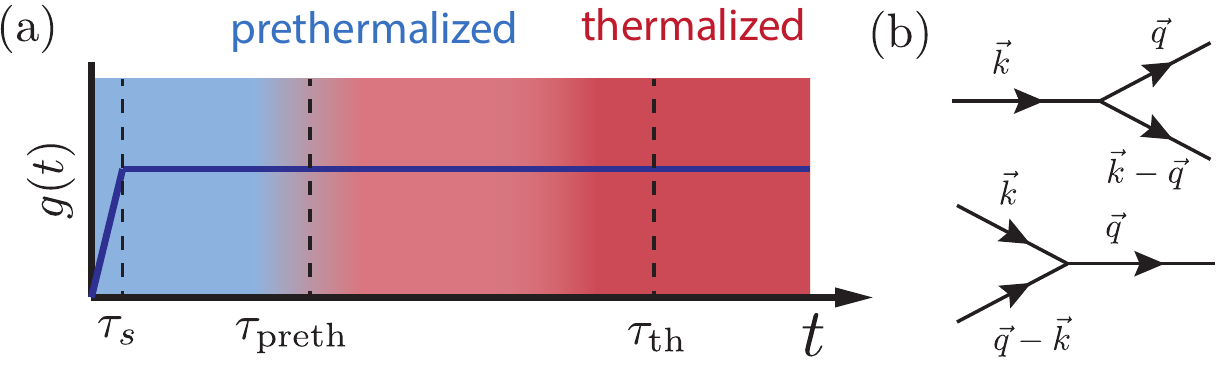}
\caption{(a) A pictorial image of the separation of time scales considered in this work. The gas is brought out of equilibrium by abruptly ramping up the interaction constant $g$ in a time $\tau_s$. Then, the approximate integrability of Hamiltonian (\ref{eq:H_momentum}) leads the system through a prethermalization stage (blue shades) on a time scale $\tau_\text{preth}$ set by the chemical potential $\mu$. Finally, a thermal equilibrium is reached on a vastly longer time scale $\tau_\text{th}$ through Beliaev-Landau collisions (red shades). Alternatively, the red shades can be seen as representing the growth of thermodynamic entropy. (b) Diagrams of the predominant non-integrable collisions that drive the system toward full thermalization; Beliaev decay (up) and Landau scattering (down).} 
\label{fig:prethermalization}
\end{figure}

Similarly, \emph{approximate} integrable systems can go through a {dephasing} stage, after which they are left in a {prethermal} state \cite{berges_prethermalization}, also described by a GGE with all the approximately conserved quantities. Nevertheless, at longer times true thermalization sets in, conveyed by higher-order relaxation processes, such as illustrated in Fig. \ref{fig:prethermalization}(a).

\hk{As of now, the literature on the crossover from a prethermalized to a thermalized state after a global quench has been mostly restricted to toy models. It has been studied how a 1D chain \cite{essler_prethermalization} or liquid \cite{diehl_prethermalization} of spinless fermions with weak integrability breaking first relaxes to a prethermal state, after which a kinetic picture allows us to understand the full thermalization dynamics of the model.
In this article, we aim to bring the study of the two distinct relaxation mechanisms within the realm of current experiments.}

\section{The Hamiltonian}

\hk{The weakly interacting Bose gas provides a natural example of an experimentally relevant system that is close to being integrable. It is}
described by the standard Hamiltonian in 3D (we use units of $\hbar=1$)
\begin{equation}
\label{eq:H_momentum}
\hat{H} = \sum_\vec{k} \frac{k^2}{2m} \, \hat{a}^\dagger_\vec{k} \hat{a}_\vec{k} + \frac{g}{2V} \sum_{\vec{p},\vec{k},\vec{q}} \hat{a}^\dagger_{\vec{p}+\vec{q}} \hat{a}^\dagger_{\vec{k}-\vec{q}}\hat{a}_\vec{k}\hat{a}_\vec{p}.
\end{equation}
Here, $V$ is the volume of the gas, $m$ is the particle mass and $g$ is the effective interaction constant, found from the $s$-wave scattering length $a_s$ as $g = 4\pi  a_s/m$. 

Starting from the ground state of an ideal gas with density $n$, we perform an abrupt ramp of the interaction constant $g_i \rightarrow g_f$ (with $g_i=0<g_f$) within a nonzero time window $\tau_s$ (see Fig. \ref{fig:prethermalization}(a)), and study the subsequent dynamics under Hamiltonian (\ref{eq:H_momentum}). Experimentally, this can be done with a Feshbach resonance \cite{review_feshbach} by suddenly ramping up an external magnetic field. Recently, this mechanism was utilized to probe the analog of cosmic Sakharov oscillations in a 2D bosonic gas \cite{chin_sakharov}. In low dimensions, the interaction constant can also be modified by varying the transverse confinement \cite{westbrook_DCE}.

When interactions are weak (small $na_s^3$) and \hk{the temperature} well below the critical temperature, almost all particles are found in the $\vec{k}=0$ mode, justifying the replacement $\hat{a}_0 \rightarrow \langle \hat{a}_0 \rangle \equiv  \sqrt{n_0 V}$, where $n_0\approx n$ is the condensate density\footnote{We perform a simplified approach, where the number of particles is not conserved, but number-conserving approaches \cite{castin_dum} would result in exactly the same Hamiltonians $\hat{H}_2$ and $\hat{H}_3$ \cite{castin_coherence}}. The dynamics of the bosonic gas after an interaction quench was studied on the level of a quadratic approximation in \hk{the} fluctuation operators \hk{$\hat{a}_{\vec{k}}$} ($\vec{k}\neq0$) \cite{natu_quench,iacopo_DCE}, and later the departure from the prethermalized state was considered \cite{menegoz_quench} and the damping of the oscillations was added by hand \cite{levin_quench}.

We, however, seek to explicitly retain terms containing three fluctuation operators as well, so as to describe higher-order (non-integrable) scatterings that eventually lead the system toward thermalization. In the literature on superfluidity, these are commonly studied in the context of Beliaev decay and Landau damping \cite{BL}, where they are responsible for the damping of a phonon \cite{pitaevskii_BL,giorgini_BL,dalibard_bl,steinhauer_damping}.

By truncating (\ref{eq:H_momentum}) to third order in fluctuation operators,
we find the approximate Hamiltonian,
\begin{equation}
\label{eq:H_bel}
\hat{H} \approx E_0 + \hat{H}_2 + \hat{H}_3.
\end{equation}
 The quadratic part can be diagonalized with the standard Bogoliubov transformation $\hat{a}_\vec{k} = u_k \hat{\chi}_\vec{k} + v_k\hat{\chi}^\dagger_{-\vec{k}}$, with 
\begin{equation}
\label{eq:bog_rot}
u_k, v_k = \pm \sqrt{ \frac{ k^2/2m +g n_0}{2\omega_k} \pm \frac{1}{2} },
\end{equation}
and the quasiparticle frequency
\begin{equation}
\label{eq:bog_spectrum}
\omega_k = \sqrt{ \frac{k^2}{2m} \Big(\frac{k^2}{2m} +2gn_0\Big)}.
\end{equation}
In terms of the Bogoliubov operators, Hamiltonian (\ref{eq:H_bel}) is then expressed as \cite{castin_coherence}
\begin{eqnarray}
\hat{H}_2 &=&  \sum_\vec{k} \omega_k \hat{\chi}_\vec{k}^\dagger\hat{\chi}_\vec{k},\\\nonumber
\hat{H}_3 &=& g\sqrt{\frac{ n_0}{V}} \sum_{\vec{k},\vec{q}}\Big( A_{\vec{k},\vec{q}}  \hat{\chi}_\vec{k}^\dagger\hat{\chi}_\vec{q}^\dagger\hat{\chi}_{\vec{k}+\vec{q}} +\\
&& B_{\vec{k},\vec{q}}\hat{\chi}_\vec{k}\hat{\chi}_\vec{q}\hat{\chi}_{-\vec{k}-\vec{q}} + \text{h.c}\; \Big), \label{eq:H3}
\end{eqnarray}
with the matrix elements of $\hat{H}_3$ 
\begin{eqnarray}
\label{eq:coefficients}
\nonumber
A_{\vec{k},\vec{q}} &=& u_\vec{k} u_\vec{q} u_{\vec{k}+\vec{q}} + v_\vec{k} v_\vec{q} v_{\vec{k}+\vec{q}}\\
&& + \big(u_{\vec{k}+\vec{q}} + v_{\vec{k}+\vec{q}}\big)\big( u_\vec{k} v_\vec{q} + u_\vec{q} v_\vec{k} \big),\\
B_{\vec{k},\vec{q}}&=&\frac{1}{3} (u_\vec{k} u_{-\vec{k}-\vec{q}} v_\vec{q} + (u_\vec{q}+v_\vec{q}) (u_\vec{k} v_{-\vec{k}-\vec{q}} + u_{-\vec{k}-\vec{q}} v_\vec{k}) \notag \\ 
&&+ 
v_\vec{k} v_{-\vec{k}-\vec{q}} u_\vec{q}).
\end{eqnarray}
Upon taking the thermodynamic limit and rescaling the wave numbers with the healing length \hk{after the quench} $\xi=\sqrt{1/m\mu}$, $k\to \tilde{k}=k\xi$ one notes that the density of states times the matrix elements of $\hat{H}_3$ squared scales as $1/(n\xi^3)=\sqrt{(4\pi)^3na_s^3}$, exactly like the condensate depletion $n-n_0$.
Therefore, if the number of depleted particles is sufficiently small, the dynamics under the integrable Hamiltonian $\hat{H}_2$ occurs on a substantially faster time scale than the ergodic dynamics of $\hat{H}_3$. \hk{A similar reasoning to compare third and 
fourth order terms of the Hamiltonian justifies the omission of $\hat{H}_4$ in \eqref{eq:H_bel}.}

\begin{figure}
\centering
\includegraphics[width = 1\columnwidth,trim = {.7cm 0 1cm 0}, clip]{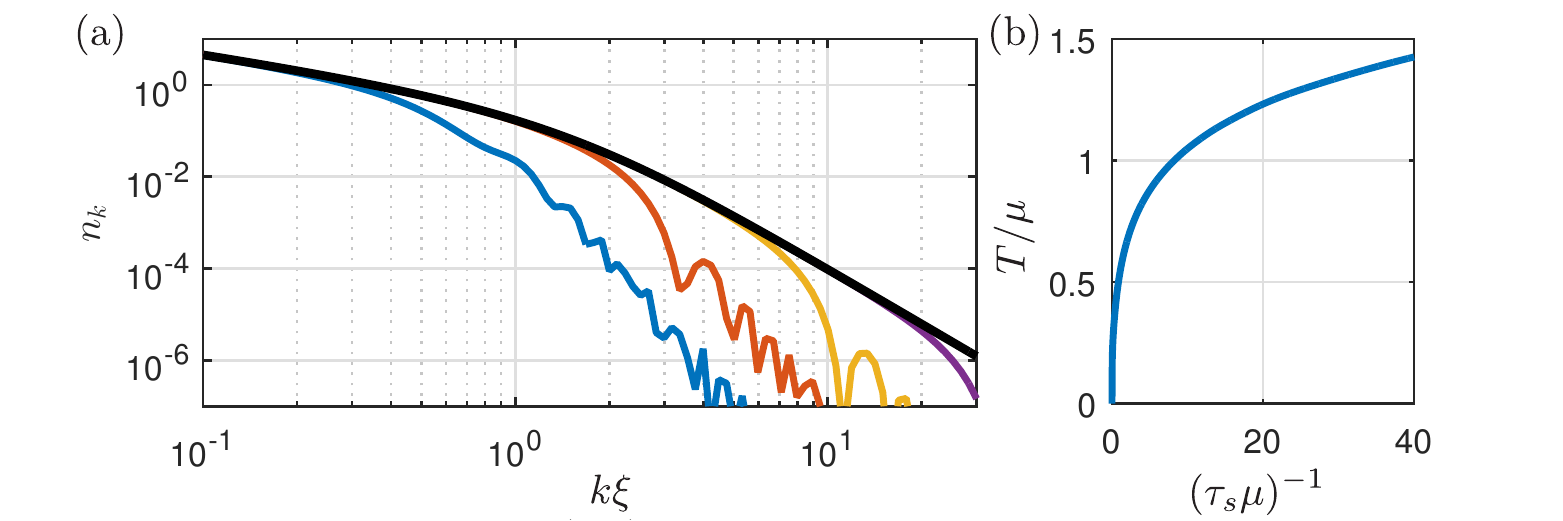}
\caption{Effect of an interaction quench $g_i=0 \rightarrow g_f$. (a) The momentum distribution $n_\vec{k}^{(\chi)}$ of quasiparticles after the interaction ramp for decreasing switching times $\tau_s = \{5,\; 0.5,\; 0.05,\; 0.005\}\times \mu^{-1}$ \hk{from bottom to top}; the thick black line is the limiting case $\tau_s\rightarrow0$, for which we have result (\ref{eq:IS}). $n^{(\chi)}_\vec{k}$ constitutes a conserved charge of $\hat{H}_2$ and only evolves under $\hat{H}_3$. (b) Upon a decrease in the switching time $\tau_s$ more energy is injected into the system, which then translates into a higher equilibrium temperature $T$. For $\tau_s \to 0$, we derive $T=O(\tau_s^{-1/5})$. }
\label{fig:DCE}
\end{figure}

\section{The equations of motion}
We start by looking at the short-time dynamics, generated by $\hat{H}_2$, such as studied in \cite{natu_quench}. In particular, the interaction ramp takes place within a nonzero time window, short enough so that we can safely neglect any effects of $\hat{H}_3$ during the quench. We return to the basis of \hk{particle} operators and find that the dynamics of the quadratic correlation functions $n^{(a)}_{\vec{k}} = \langle \hat{a}_\vec{k}^\dagger \hat{a}_\vec{k} \rangle$ and $c^{(a)}_{\vec{k}} = \langle \hat{a}_\vec{k} \hat{a}_{-\vec{k}} \rangle$ is governed by \cite{natu_quench}
\begin{eqnarray}
\label{eq:DCE_corrs}
\!\!\!\!\partial_t n_{\vec{k}}^{(a)} &=&  -2\Im \big[g(t) n_0 c^{(a)}_{\vec{k}}\big],\\
\!\!\!\!i\partial_t c_{\vec{k}}^{(a)} &=& 2\Big(\frac{k^2}{2m} +g(t) n_0\Big) c_{\vec{k}}^{(a)} +
g(t) n_0 (2n_{\vec{k}}^{(a)} + 1).
\end{eqnarray}
This system of equations is readily integrated numerically for a given temporal profile $g(t)$ and with appropriate initial conditions. It has been intensely studied in the context of the dynamical Casimir effect \cite{DCE_review}, where a modulation of the interaction constant or condensate density causes a change of vacuum for the quasiparticle operators $\hat{\chi}_\vec{k}$ \cite{westbrook_DCE,iacopo_DCE,selma_DCE,wilson_DCE}. The correlations of the quasiparticles, in turn, are evaluated with the linear transform
\begin{eqnarray}
\label{eq:IS_trans_n}
n_{\vec{k}}^{(\chi)} &=& \big( u_k^2 + v_k^2\big) n_{\vec{k}}^{(a)} - 2u_k v_k \Re\big[c^{(a)}_{\vec{k}}\big]  +  v_k^2,\\\label{eq:IS_trans_c}
c_{\vec{k}}^{(\chi)} &=& u_k^2 c_{\vec{k}}^{(a)} + v_k^2 c_{\vec{k}}^{(a)\ast} - 2u_k v_k n_{\vec{k}}^{(a)} - u_k v_k.
\end{eqnarray}
 In the limit of instantaneous switching time $\tau_s \rightarrow0$, \hk{$n_{\vec{k}}^{(a)}$ and $c_{\vec{k}}^{(a)}$ are zero just after the quench}
 and we find the correlation functions after the quench as
\begin{equation}
\label{eq:IS}
n_{\vec{k}}^{(\chi)}  = v^2_k, \;\;\; c_{\vec{k}}^{(\chi)} = -u_kv_k.
\end{equation}
In Fig. \ref{fig:DCE}(a), we show the quasiparticle momentum distribution for different $\tau_s$ and see that it converges to (\ref{eq:IS}) for shorter $\tau_s$.

We now stick to the basis of Bogoliubov operators $\hat{\chi}_\vec{k}$. Their quadratic correlation functions evolve trivially under $\hat{H}_2$ as $n_{\vec{k}}^{(\chi)}(t) = n_{\vec{k}}^{(\chi)}$ and $c_{\vec{k}}^{(\chi)}(t) = \tilde{c}_{\vec{k}}^{(\chi)}e^{-2i\omega_k t}$, making $n_{\vec{k}}^{(\chi)}$ and $\tilde{c}_{\vec{k}}^{(\chi)}$ conserved quantities of $\hat{H}_2$ related to the integrable dynamics. However, they do experience a variation under the full Hamiltonian (\ref{eq:H_bel}), \hk{which breaks the integrability}. Via Heisenberg's equation of motion, we derive their dynamics  \hk{under  $\hat{H}_3$}:
\begin{eqnarray}
\label{eq:n_HOC}
\nonumber
\partial_t n_\vec{k}^{(\chi)} &=& 2g\sqrt{\frac{ n_0}{V}} \Im \bigg\{ \sum_\vec{q} 3B_{\vec{k},-\vec{q}} R^\ast_{\vec{k},\vec{q}}\\
&& +2A_{\vec{k},\vec{q}-\vec{k}} M_{\vec{q},\vec{k}} + A_{\vec{q},\vec{k}-\vec{q}} M^\ast_{\vec{k},\vec{q}} \bigg\} \\\nonumber
i\partial_t \tilde{c}_\vec{k}^{(\chi)} &=& 2g\sqrt{\frac{ n_0}{V}}\sum_\vec{q} \bigg\{ 3B_{-\vec{k},\vec{q}} M_{\vec{k},\vec{q}}\\
&& + 2 A_{\vec{k},-\vec{q}} M^\ast_{\vec{q},\vec{k}} + A_{\vec{q},\vec{k}-\vec{q}} R_{\vec{k},\vec{q}} \bigg\}e^{2i\omega_k t},\label{eq:c_HOC}
\end{eqnarray} 
where we have introduced the correlation functions of three quasiparticles
\begin{equation}
\label{eq:third_beliaev}
M_{\vec{k},\vec{q}} = \Big\langle \hat{\chi}^\dagger_{\vec{k}-\vec{q}} \hat{\chi}^\dagger_{\vec{q}} \hat{\chi}_{\vec{k}} \Big\rangle,\;\;\;
R_{\vec{k},\vec{q}} = \Big\langle \hat{\chi}_{\vec{q}-\vec{k}} \hat{\chi}_{-\vec{q}} \hat{\chi}_{\vec{k}} \Big\rangle.
\end{equation}
We next evaluate the equation of motion for these third-order correlators
\begin{eqnarray}
\label{eq:M_HOC}
i \partial_t M_{\vec{k},\vec{q}} &=& (\omega_\vec{k} - \omega_\vec{q} - \omega_{\vec{k}-\vec{q}}) M_{\vec{k},\vec{q}} + g\sqrt{\frac{ n_0}{V}} F^{(M)}_{\vec{k},\vec{q}},\\\label{eq:R_HOC}
i \partial_t R_{\vec{k},\vec{q}} &=& (\omega_\vec{k} + \omega_\vec{q} + \omega_{\vec{q}-\vec{k}}) R_{\vec{k},\vec{q}} + g\sqrt{\frac{ n_0}{V}} F^{(R)}_{\vec{k},\vec{q}}.
\end{eqnarray} 
\hk{Here, the matrices $F^{(M,R)}_{\vec{k},\vec{q}}$ contain correlators of four operators.} More generally, a connected correlator of $p$ operators couples to correlators of $p+1$ operators on the right-hand side, making this an ever-growing hierarchy \cite{proukakis_hierarchy}. \hk{However, as explained in \cite{beliaev_us}, fourth-order correlators in $F^{(M,R)}_{\vec{k},\vec{q}}$ can be approximately factorized into products of second-order correlators using Wick's theorem\footnote{\hk{Since the average value of linear operators $\langle\hat{\chi}_{\vec{k}}\rangle$ is zero to zeroth order in $\hat{H}_3$, the products of a first and third order correlators are negligible compared to the terms in \eqref{FMkq} and \eqref{FRkq}.}}, thus establishing a truncated hierarchy of correlations functions.}  After this factorization, we find the drive term in Eqs. (\ref{eq:M_HOC})--(\ref{eq:R_HOC}) as \hk{(we drop the superscript ${\boldsymbol{\cdot}} ^{(\chi)}$ for ease of notation)}
\begin{eqnarray}
\label{eq:F_M2}
\nonumber
F^{(M)}_{\vec{k},\vec{q}} &=& 2A_{\vec{k},-\vec{q}}\Big( c^\ast_\vec{q}( n_{\vec{k}-\vec{q}} - n_\vec{k} ) - c^\ast_{\vec{k} - \vec{q}} c_\vec{k} \Big)\\\nonumber
&& + 2A_{ \vec{k},\vec{q}-\vec{k}} \Big( c^\ast_{\vec{k}-\vec{q}}(n_\vec{q} - n_\vec{k} ) - c_\vec{k} c^\ast_\vec{q} \Big)\\\nonumber
&&  + 2A_{ \vec{q},\vec{k}-\vec{q}} \Big( n_{\vec{k}-\vec{q}}( n_\vec{q} - n_\vec{k} ) - n_\vec{k}(n_\vec{q} +1) \Big)\\\nonumber
&& + 3B_{\vec{k},-\vec{q}} \Big(  c^\ast_{\vec{k} - \vec{q}}c^\ast_\vec{q} - c_\vec{k} n_{\vec{k}-\vec{q}} \Big)\\\nonumber
&& + 3B_{\vec{k},\vec{q}-\vec{k}}\Big( c^\ast_{\vec{k}-\vec{q}} c^\ast_\vec{q} - c_\vec{k}(n_{\vec{q}}  + 1)\Big)\\
&& - 3B_{\vec{q},\vec{k}-\vec{q}}\;c_\vec{k}\Big(n_\vec{q} +  n_{\vec{k}-\vec{q}} +1 \Big), \label{FMkq}
\end{eqnarray}
and 
\begin{eqnarray}
\label{eq:F_R2}
\nonumber
F^{(R)}_{\vec{k},\vec{q}} &=& 2 A_{\vec{k},-\vec{q}}\Big( c_{\vec{k}-\vec{q}}( n_\vec{k} + n_\vec{q} + 1 ) + c_\vec{k}c_\vec{q} \Big)\\\nonumber
&& + 2A_{\vec{k},\vec{q}-\vec{k}}\Big( c_\vec{q}( n_\vec{k} + n_{\vec{k}-\vec{q}} + 1 ) + c_{\vec{k}-\vec{q}}c_\vec{k} \Big)\\\nonumber
\label{eq:F_2w}
&& + 2A_{\vec{q},\vec{k}-\vec{q}}\Big( c_\vec{k}( n_\vec{q} + n_{\vec{k}-\vec{q}} + 1) + c_\vec{q}c_{\vec{k}-\vec{q}} \Big)\\\nonumber
&& + 3B_{\vec{k},-\vec{q}}\;\Big( (n_{\vec{k}-\vec{q}} + 1)(n_\vec{k} + n_\vec{q} + 1) \Big)\\\nonumber
&& + 3B_{\vec{k},\vec{q}-\vec{k}}\Big( n_\vec{q}(n_\vec{k} + n_{\vec{k}-\vec{q}} + 1) +  n_{\vec{k}-\vec{q}} + 1\Big)\\
&& + 3B_{\vec{q},\vec{k}-\vec{q}}\; \Big( n_\vec{k}( n_\vec{q} + n_{\vec{k}-\vec{q}} +1 ) \Big). \label{FRkq}
\end{eqnarray}
\hk{As such, we establish a closed set of differential equations for correlators up to order three, which approximately describes the dynamics of the bosonic gas after the interaction ramp, provided $n_0 \gg n-n_0$, ensuring that connected correlators of higher order have a decreasing magnitude.}

\section{The kinetic equations}
\begin{figure}
\centering
\includegraphics[scale=.55, trim={ 0cm 0 1cm 0 }, clip]{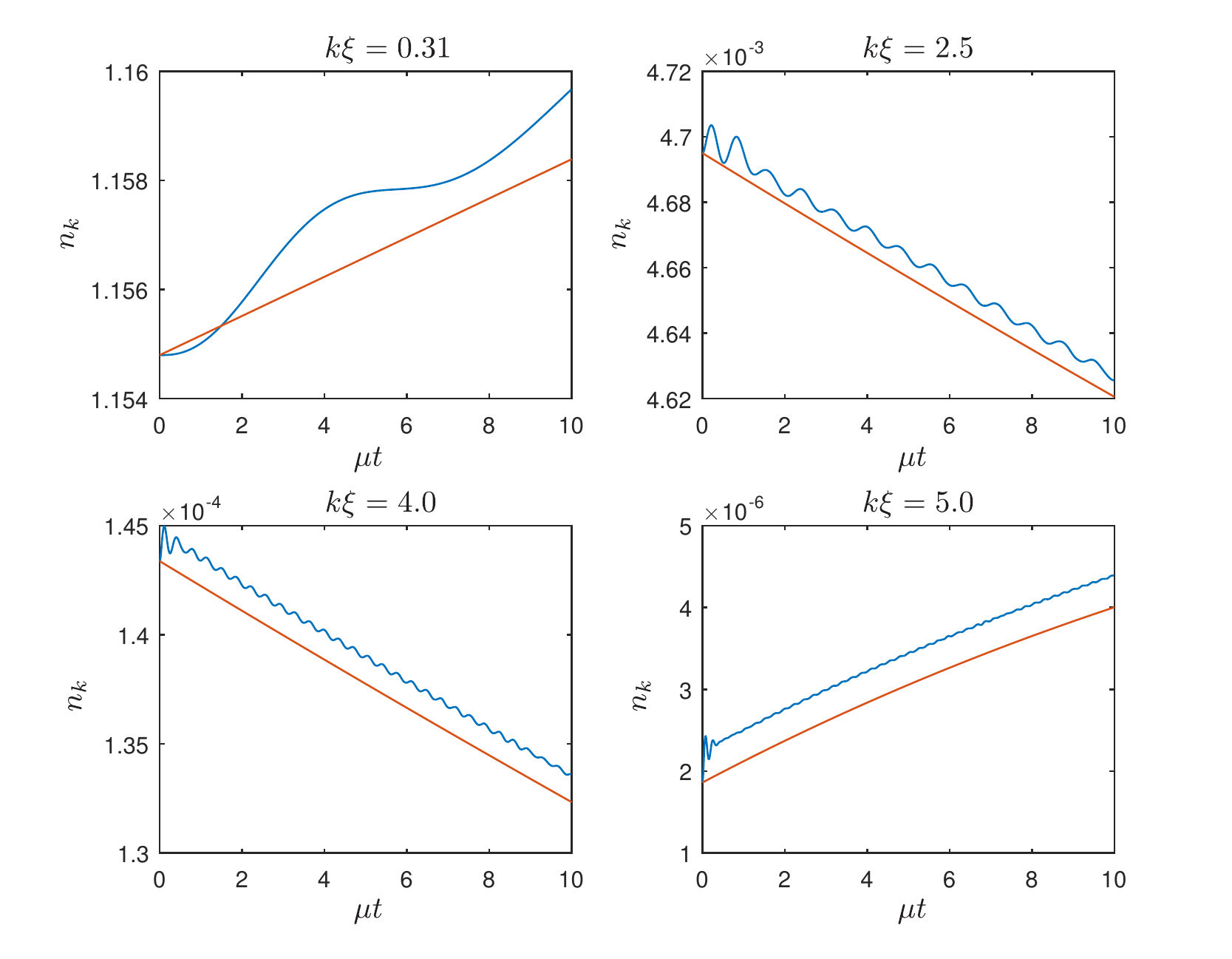}
\caption{A comparison of the quasiparticle occupation numbers $n_\vec{k}$ as produced by the integration of the hierarchy of correlation functions (blue lines) and the derived kinetic equation in the adiabatic limit (red lines) for the quench $g_i=0 \rightarrow g_f = 0.05\mu\xi^3$ (corresponding to $n\xi^3=20$) in $\tau_s = 0.5\mu^{-1}$ for different momenta (see the initial momentum distribution of quasi-particles right after the quench on Fig. \ref{fig:DCE}). }
\label{fig:compare}
\end{figure}

In the long-time limit, the coupled system of equations (\ref{eq:n_HOC})--(\ref{eq:c_HOC}) and (\ref{eq:M_HOC})--(\ref{eq:R_HOC}) reproduces the well-known kinetic equations. This can be seen by formally solving (\ref{eq:M_HOC}) as
\begin{equation}
\label{eq:M_sol}
M_{\vec{k},\vec{q}}(t) = -ig\sqrt{\frac{n_0}{V}}\int_{0}^{t} ds\;F_{\vec{k},\vec{q}}^{(M)}(s)\; e^{i(\omega_\vec{k} -\omega_\vec{q}-\omega_{\vec{k}-\vec{q}})(s-t)},
\end{equation}
and similar for $R_{\vec{k},\vec{q}}(t)$ in (\ref{eq:R_HOC}). These expressions can now be plugged into (\ref{eq:n_HOC})--(\ref{eq:c_HOC}), after which we obtain effective dynamics by $(i)$ sending the integration boundary $t\rightarrow \infty$ in (\ref{eq:M_sol}), thus singling out non-oscillating terms in the \hk{integral over $s$}, and $(ii)$ time averaging Eq.~\eqref{eq:n_HOC} \hk{to remove the contributions that oscillate rapidly with time $t$}. The result is that the evolution of quasiparticle occupation numbers is governed by the kinetic (or quantum Boltzmann) equations 
\begin{eqnarray}
\label{eq:n_kin}
\nonumber
\partial_t n_\vec{k} &=& 4\pi \frac{ g^2 n_0}{V} \bigg\{\sum_\vec{q} A^2_{\vec{q},\vec{k}-\vec{q}} \delta\big( \omega_\vec{k} - \omega_\vec{q} - \omega_{\vec{k}-\vec{q}}\big) \\\nonumber &&
\;\;\;\times\Big( n_{\vec{k}-\vec{q}} n_\vec{q} - n_\vec{k}( n_\vec{q} + n_{\vec{k}-\vec{q}} + 1 )\Big)  \\\nonumber
&& + 2\sum_\vec{q} A^2_{\vec{k},\vec{q}-\vec{k}} \delta\big( \omega_\vec{q} - \omega_\vec{k} - \omega_{\hk{\vec{q}-\vec{k}}}\big) \\
&& \;\;\;\times \Big( n_\vec{q}(n_\vec{k} + n_{\hk{\vec{q}-\vec{k}}} + 1) - n_\vec{k} n_{\hk{\vec{q}-\vec{k}}}\Big) \bigg\}.
\end{eqnarray}
Within the \hk{kinetic} approximation, the oscillation frequencies from the evolution of $M_{\vec{k},\vec{q}}$ have been translated into $\delta$-functions imposing energy conservation for the redistribution of quasiparticle occupation numbers.  In our method, the kinetic equations come as a limiting behavior, so that deviations from them can be studied quantitatively, \hk{as we do in Fig.~\ref{fig:compare}}; this to our knowledge has not been done previously in 3D.

With (\ref{eq:n_kin}), we rederive the kinetic equation that is known from the literature on Beliaev-Landau scattering, where it is commonly established with Fermi's golden rule \cite{castin_coherence}. The first term represents the redistribution of quasiparticles through Beliaev decay, where a quasiparticle with high momentum $\vec{k}$ decays into (or is formed from) two with $\vec{q}$ and $\vec{k}-\vec{q}$. The second term, in turn, describes the \hk{Landau process of} absorption (or emission) of the quasiparticle with momentum $\vec{k}$ by a quasiparticle $\vec{q}-\vec{k}$ (or $\vec{q}$). Notice that the \hk{Landau} term comes with an additional factor $2$ from the two possible scattering channels \cite{giorgini_BL}. See Fig. \ref{fig:prethermalization}(b) for the corresponding diagrams.

Through the same analysis, we obtain the evolution of pair correlations,
\begin{equation}
i\partial_t {c}_\vec{k}=\bb{2\omega_\kk+2\delta\omega_\kk-i{\gamma_\kk}}{c}_\vec{k} +I_\kk(\{c_\qq\})
\label{ck}
\end{equation}
The first term ur this equation contains the evolution of ${c}_\vec{k}$ under the Bogoliubov frequency $z_\kk=\omega_\kk+\delta\omega_\kk-i{\gamma_\kk}/2$
calculated treating $\hat{H}_3$ in second order perturbation theory in the instantaneous Fock state $|n_\kk,\{n_\qq\}_{\qq\neq\kk}\rangle$ (see the Appendix for the explicit
derivation). This contains the Landau-Beliaev damping rate
\begin{widetext}
\begin{equation}
\label{eq:gamBL}
\gamma_\vec{k} =4\pi \frac{ g^2 n_0}{V} \sum_\qq  \bbcro{A^2_{\vec{q},\vec{k}-\vec{q}} ( n_\vec{q} + n_{\vec{k}-\vec{q}} + 1 ) \delta\big( \omega_\vec{k} - \omega_\vec{q} - \omega_{\vec{k}-\vec{q}}\big) 
+ 2 A^2_{\vec{k},\vec{q}-\vec{k}} (n_{\vec{q}-\vec{k}}-n_\vec{q})\delta\big( \omega_\vec{q} - \omega_\vec{k} - \omega_{\vec{q}-\vec{k}}\big),}
\end{equation}
(where the Beliaev and Landau parts are respectively the first and second summation) and the frequency shift
\begin{equation}
\label{domega}
\delta\omega_\vec{k} = \frac{ g^2 n_0}{V} \mathcal{P} \sum_\qq \Bigg[\frac{2A^2_{\vec{q},\vec{k}-\vec{q}} ( n_\vec{q} + n_{\vec{k}-\vec{q}} + 1 )} { \omega_\vec{k} - \omega_\vec{q} - \omega_{\vec{k}-\vec{q}}} 
+\frac{4A^2_{\vec{k},\vec{q}-\vec{k}} (n_{\vec{q}-\vec{k}}-n_\vec{q})} { \omega_\vec{k} + \omega_{\vec{q}-\vec{k}}-\omega_\vec{q} }
-\frac{18B^2_{\vec{k},\vec{q}} ( n_\vec{q} + n_{-\vec{k}-\vec{q}} + 1 )} { \omega_\vec{k} + \omega_{\vec{q}}+\omega_{-\vec{k}-\vec{q}} }\Bigg],
\end{equation}
where $\mathcal{P}$ denotes the Cauchy principal value and we have used the fact that our momentum distribution remains symmetric $n_\kk^{(\chi)}=n_{-\kk}^{(\chi)}$.
The collisional integral $I_\kk(\{c_\qq\})$ accounts for the fact that the distribution $\{c_\qq\}$ in modes $\qq\neq\kk$ changes
dynamically with $c_\kk$; it is given by
\begin{equation}
\label{Ik}
I_\kk(\{c_\qq\}) =  2\frac{ g^2 n_0}{V} \Bigg[ 
\sum_\vec{q} 
  \frac{2A^2_{\vec{q},\vec{k}-\vec{q}} {c}_\vec{q} {c}_{\vec{k}-\vec{q}} }{\omega_\qq+\omega_{\kk-\qq}-\omega_\kk+i0^+}   
+ \frac{4 A^2_{\vec{k},\vec{q}-\vec{k}}{c}_\vec{q} {c}^\ast_{\vec{q}-\vec{k}}}{\omega_{\qq} - \omega_{\kk} - \omega_{\qq-\kk}+i0^+}  
+\frac{18B_{\kk,\qq}^2\tilde{c}_\vec{-\kk-\qq}^{*} \tilde{c}_\vec{\qq}^{*} }{-\omega_\kk-\omega_{\qq}-\omega_{-\kk-\qq}+i0^+}  \Bigg],
\end{equation}
Eq.~\eqref{ck} describes how the coherence between modes $\kk$ and $-\kk$ evolves under
three-body scatterings, and in particular $I_\kk$ describes how it is affected by the coherence in other modes
$\qq$. Remark that $c_\kk$ may show a temporal evolution even if the populations $n_\kk$ are prepared at thermal equilibrium, reflecting the underlying Landau-Beliaev scatterings
that maintain equilibrium. 
To our knowledge, this equation was not found in the literature, unlike the kinetic equation \eqref{eq:n_kin} on $n_\kk$ \cite{LandauLipschitzVol10}.
\end{widetext}

\hk{In Fig.~\ref{fig:compare}, we perform} a quantitative comparison between the full integration of the truncated hierarchy and the approximate kinetic description (\ref{eq:n_kin}). We show the evolution of the quasiparticle occupation numbers at short times $t\sim1/\mu$. We observe that the curve of $n_k(t)$ predicted by the kinetic equation differs in two distinct ways from that of the hierarchy: $(i)$ the evolution at very short times is not well captured by the kinetic equation, which results in a small offset (controlled by the interaction strength $na_s^3$) between the two curves, \hk{an offset then} conserved all along the evolution and $(ii)$ contrary to the kinetic description, the hierarchy of correlations retains high-frequency components in $n_k(t)$. Those two differences are directly related to the approximations $(i)$ and $(ii)$ detailed in the main text below \eqref{eq:M_sol}, on which kinetic equations are based.

In the long-time limit, we find that (\ref{eq:n_kin}) and (\ref{ck}) converge to the values in a thermal ensemble. The momentum distribution of quasiparticles approaches the Bose-Einstein distribution
\begin{equation}
\label{eq:n_th}
n^\text{th}_\vec{k} = \frac{1}{e^{\beta \omega_\vec{k}}-1},
\end{equation}
with $\beta = 1/k_B T$ the inverse temperature set by the total injected energy, while the anomalous correlations vanish. The energy after the quench on the level of the quadratic Hamiltonian, $E=E_0+\sum_\vec{k}\omega_k \,n^{(\chi)}_k $, is conserved under the kinetic equations. \hk{However, using the value of the mode occupation number $n^{(\chi)}_k$ for an infinitely fast quench [see Eq.~(\ref{eq:IS})] leads to  an ultraviolet divergence of this injected energy. This divergence is regularized by a finite switching time: this sets an effective cutoff in energy $\omega_{k_{\rm max}}\propto1/\tau_s$, corresponding to a momentum cutoff $k_{\rm max}\propto1/\sqrt{\tau_s}$ in the limit of fast quench $\tau_s\to0$, and therefore to an injected energy $E-E_0 \propto 1/\sqrt{\tau_s}$}. This enables us to fix the total injected energy with the switching time $\tau_s$ and, consequently, the final equilibrium temperature of the gas by matching this energy with the energy of a thermal ensemble. When $k_BT>\mu$, we have that $E-E_0\propto T^{5/2}$ , such that we derive the asymptotic scaling $T\propto \tau_s^{-1/5}$. In Fig. \ref{fig:DCE}(b), we show the full variation of equilibrium temperature with switching time $\tau_s$.

\section{The density-density correlation function}
Finally, we investigate the behavior of macroscopic observables in real space, which are expected to exhibit the two distinct relaxation stages. We concentrate on distances of the order of the \hk{(equilibrium)} thermal wavelength $2\pi/k_\text{th}$, with $\omega_{k_\text{th}} = k_BT$. We choose $\tau_s^{-1}= 0.5\mu$, such that $k_BT=0.67\mu$ and therefore the thermal wavelength is of the same order as the healing length $\xi$. For $k\sim1/\xi$, \hk{the kinetic equations are accurate \hk{for times $t>1/\mu$}, 
and introduce an offset of the order of $\sqrt{na_{\rm s}^3}$, our small parameter. Therefore, they correctly describe the dynamics of
spatial correlations at length scales $\sim\xi$ in the weakly interacting limit, as we have also checked numerically.}

\begin{figure}
\centering
\includegraphics[scale=.65, trim = {.6cm 0 .5cm 0}, clip]{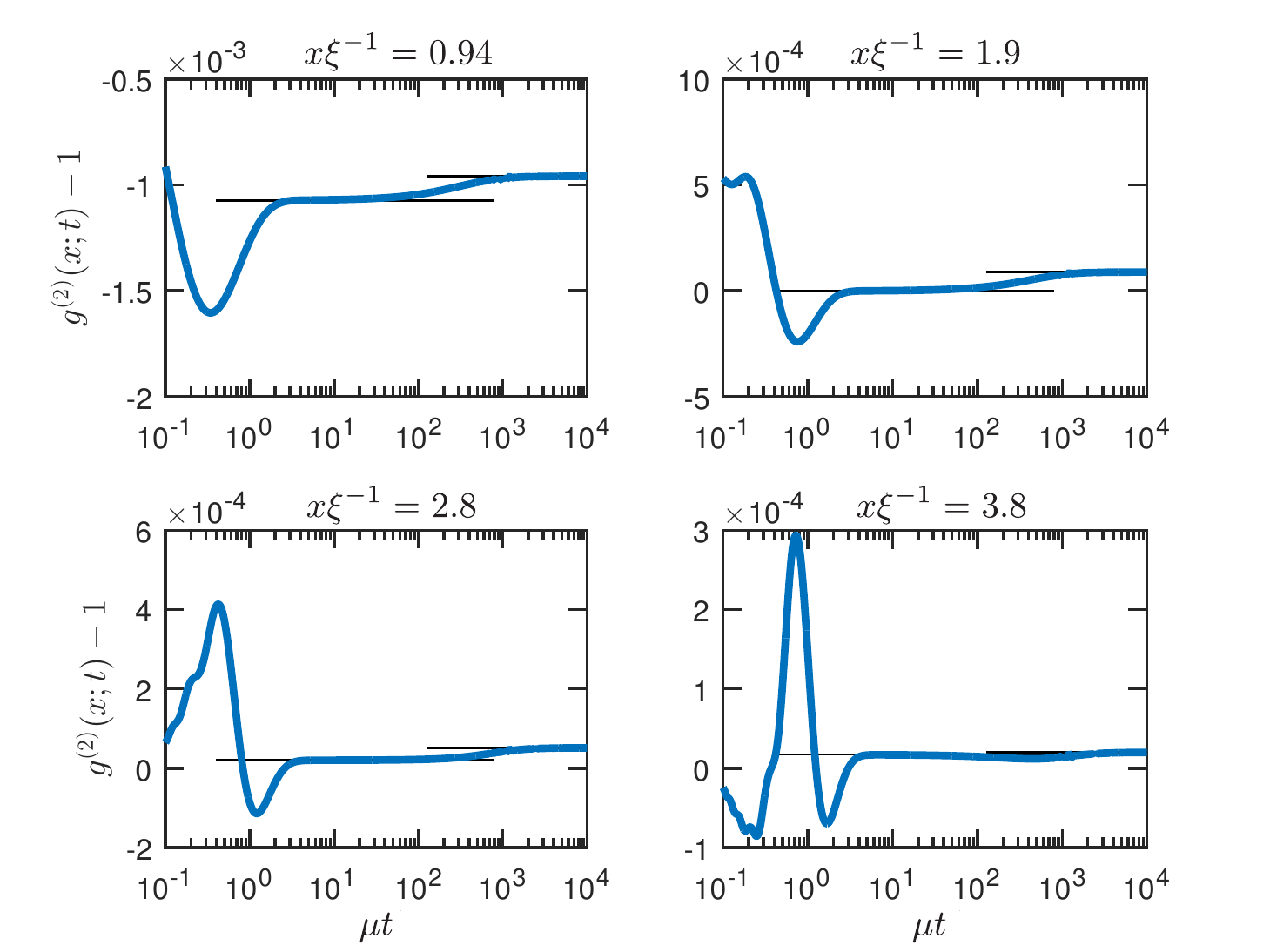}
\caption{The evolution of the density correlation function after the quench $g_i=0\rightarrow g_f=0.05\mu\xi^3$ in $\tau_s=0.5\mu^{-1}$ for varying distances $x=|\vec{r}-\vec{r}'|$ \hk{according to the kinetic picture Eqs.~\eqref{eq:n_kin} and \eqref{ck}}. The \hk{horizontal solid} lines indicate the asymptotic values for the prethermal and thermal (quasi)stationary ensemble; the temperature $T=0.67\mu$ is found from the initial state. At short distances, we clearly notice a relaxation to a prethermal plateau on a time scale of the order of $\tau_\text{preth}=\mu^{-1}$ (for ramps $\tau_s\sim \mu^{-1}$ and $x\sim \xi$), this is due to a dephasing mechanism in $\hat{H}_2$. In $g^{(2)}(x;t)$, this is manifested as a fast oscillation at short times, which then diminishes due to a destructive interference between all $\vec{k}$-modes once the light-cone correlation peak has moved away from the considered 
distance $x$ \cite{iacopo_DCE}. Then, at much later times, $\tau_\text{therm}\sim10^3\mu^{-1}$, a new equilibrium value is found that corresponds to the value in the thermal ensemble through the much slower dynamics of $\hat{H}_3$. The difference between the prethermal and thermal value vanishes for increasing separation $x$ as the correlation function drops to zero in this limit.  }
\label{fig:g2_plot}
\end{figure}
The first relaxation stage of local observables to their prethermal value is caused by a dephasing mechanism where all $\vec{k}$-modes interfere desructively. We therefore define the annihilation operator in position space $\hat{a}(\vec{r}) = 1/\sqrt{V}\sum_\vec{k}e^{i\vec{k}\cdot\vec{r}} \hat{a}_\vec{k}$. Our analysis is now focused on the evolution of the density-density correlation function, defined as $g^{(2)}(\vec{r}-\vec{r}';t) = \langle : \hat{n}(\vec{r})\,\hat{n}(\vec{r}'):\rangle_t/\langle \hat{n}(\vec{r}) \rangle_t^2$ for a homogeneously distributed gas, where `$:$' denotes normal ordering and $\hat{n}(\vec{r}) = \hat{a}^\dagger(\vec{r})\hat{a}(\vec{r})$ is the local density operator. The density-density correlation function has proven its importance previously in the context of analog gravity \cite{iacopo_hawking,iacopo_hawking2}, where the correlation pattern shows a fingerprint of the analog of Hawking radiation at an acoustic black hole's horizon \cite{steinhauer_hawking}. 

On the Gaussian level, the density correlation function can be simplified to 
\begin{equation}
g^{(2)}(\vec{r}-\vec{r}';t) = 1 + \frac{2}{n_0}\big( \mathfrak{n}(\vec{r}-\vec{r}';t) + \text{Re}\big\{ \mathfrak{m}(\vec{r}-\vec{r}';t)\big\}\big),
\end{equation}
where we defined
\begin{equation}
\mathfrak{n}(\vec{r}-\vec{r}';t) = \langle \hat{a}^\dagger(\vec{r}) \hat{a}(\vec{r}')\rangle_t=\frac{1}{V} \sum_{\vec{k}\neq0} e^{i\vec{k}\cdot(\vec{r}-\vec{r}')} n_{\vec{k}}^{(a)}(t),
\end{equation}
and analogous for $\mathfrak{m}(\vec{r}-\vec{r}') = \langle \hat{a}(\vec{r}) \hat{a}(\vec{r}')\rangle$. The quadratic correlations of fluctuations, $n_{\vec{k}}^{(a)}$ and $c_{\vec{k}}^{(a)}$, can be obtained from the quasiparticle correlations $n_\vec{k}^{(\chi)}$ and $c_\vec{k}^{(\chi)}$ through the inverse of the transformation (\ref{eq:IS_trans_n})--(\ref{eq:IS_trans_c}).

In Fig. \ref{fig:g2_plot} we show the evolution of the density correlation function after a ramp $g_i=0\rightarrow g_f=0.05\mu \xi^3$ (so that $na^3_s=1.3\cdot10^{-5}$) at different distances $x=|\vec{r}-\vec{r}'|$. We observe a clear first relaxation, approximately to the prethermal value on a time scale $\tau_\text{preth}\sim\mu^{-1}$ after an initial oscillation due to the light-cone peak \cite{calabrese_qp,iacopo_DCE} that dies out due to dephasing once this has traveled away; this is governed by $\hat{H}_2$. At much longer times, the scatterings contained in $\hat{H}_3$ cause a new relaxation, this time to the thermal value. We find that the thermalization time $\tau_\text{therm}\sim 10^3\mu^{-1}$ is in qualitative agreement with the Beliaev-Landau lifetime of the thermal wavenumber $1/\gamma^{BL}_{k_\text{th}}\sim1/\mu\sqrt{na_s^3}$ for $k_BT\sim\mu$ \cite{castin_coherence}.

\section{Conclusions}
We have illustrated that the crossover from a prethermalized to a thermalized state can be witnessed in a cold atomic gas by probing the density correlations after a sudden interaction ramp. The switching time of the ramp determines the final temperature in the equilibrium ensemble. While a simple dephasing mechanism, treated on the level of the \hk{(integrable)} quadratic Hamiltonian, causes local observables to relax to a prethermal value, a more sophisticated approach is needed to describe the thermalization stage. Here, third-order interaction processes, known as Beliaev-Landau collisions, are the predominant mechanism to lead the system away from integrability and, \hk{eventually}, to thermal equilibrium. When focusing on \hk{most relevant} length scales of the order of the \hk{equilibrium} thermal wavelength, a kinetic description is sufficient to describe the final relaxation. In principle, our predictions are within reach of current experiments with ultracold atomic gases.

\acknowledgments
MVR gratefully acknowledges support in the form of a Ph. D. fellowship of the Research Foundation - Flanders (FWO) and hospitality at the BEC Center in Trento.  HK is supported by the FWO and the European Union H2020 program under the MSC Grant Agreement No. 665501. MW acknowledge financial support from the FWO-Odysseus program.
IC was funded by the EU-FET Proactive grant AQuS, Project No. 640800, and by Provincia Autonoma di Trento, partially through the project ``On silicon chip quantum optics for quantum computing and secure communications (SiQuro)''.

\appendix

\section*{Appendix: Perturbed Bogoliubov energy in an arbitrary excited state}

\hk{To recover Eqs.~(\ref{eq:gamBL}--\ref{domega}), we treat $\hat{H}_3$ as a perturbation
of $\hat{H}_2$ and we recall \cite{Cohen,Annalen} that the complex poles $E$ of the resolvent
(or equivalently of the Green function) in a given state $|\psi\rangle$
 are given to second order in the perturbation by
\begin{eqnarray}
E&=&E_0+\langle\psi|\hat{H}_3|\psi\rangle+\langle\psi|\hat{H}_3\frac{\hat{Q}}{E-\hat{H}_2}\hat{H}_3|\psi\rangle\\
&=& E_0+\langle\psi|\hat{H}_3|\psi\rangle+\sum_{\lambda}\frac{|\langle\psi|\hat{H}_3|\lambda\rangle|^2}{E-\langle\lambda|\hat{H}_2|\lambda\rangle}
\label{PGP}
\end{eqnarray}
where $E_0=\langle\psi|\hat{H}_2|\psi\rangle$ is the unperturbed energy, $\hat{Q}=1-|\psi\rangle\langle\psi|$
projects orthogonally to $|\psi\rangle$ and the states $|\lambda\rangle$ are therefore orthogonal to $|\psi\rangle$. 
We apply Eq.~\eqref{PGP} to the Fock states $|n_\kk,\{n_\qq\}_{\qq\neq\kk}\rangle$ 
and $|n_\kk-1,\{n_\qq\}_{\qq\neq\kk}\rangle$ whose perturbed energies, respectively $E(n_\kk)$ and $E(n_\kk-1)$,
define the perturbed Bogoliubov frequency
\begin{equation}
z_\kk\equiv E(n_\kk)-E(n_\kk-1)
\end{equation}
Changing the sum over the intermediate Fock states $\lambda$ 
into a sum over the scattered momentum $\qq$ (taking care to avoid double 
countings) and replacing in the denominator $E(n_\kk)$ by its zeroth-order approximation
$E_0(n_\kk)$ we get
\begin{multline}
\!\!\!\!\!\!z_\kk\!=\!z_{\kk,0}\\+g^2\frac{n_0}{V}\sum_\qq \!\Bigg[\! \frac{2A_{\qq,\kk-\qq}^2 [(1+n_{\kk-\qq})(1+n_\qq)-n_{\kk-\qq}n_\qq]}{z_{\kk,0}-\omega_{\qq}-\omega_{\kk-\qq}} \\
+\frac{4A_{\kk,\qq-\kk}^2 [(1+n_{\qq})n_{\qq-\kk}-n_{\qq}(1+n_{\qq-\kk})]}{z_{\kk,0}+\omega_{\qq-\kk}-\omega_{\qq}} \\
+\frac{18B_{\kk,\qq}^2 [n_{-\kk-\qq}n_\qq-(1+n_{-\kk-\qq})(1+n_\qq)]}{z_{\kk,0}+\omega_{\qq}+\omega_{-\kk-\qq}} \Bigg]
\end{multline}
where $z_{\kk,0}=E_0(n_\kk)-E_0(n_\kk-1)=\omega_\kk+i0^+$ is the unperturbed Bogoliubov
frequency. The counting factors originate in our symmetric writing \eqref{eq:H3} of $\hat{H}_3$
where the same operator appears more than once. Using the Plemelj formula $1/(x+i0^+)=\mathcal{P}(1/x)
-i\pi\delta(x)$ to split the real and imaginary parts finally yields Eqs.~(\ref{eq:gamBL}--\ref{domega}).}
 \bibliography{bibliography}

\begin{thebibliography}{37}%
\makeatletter
\providecommand \@ifxundefined [1]{%
 \@ifx{#1\undefined}
}%
\providecommand \@ifnum [1]{%
 \ifnum #1\expandafter \@firstoftwo
 \else \expandafter \@secondoftwo
 \fi
}%
\providecommand \@ifx [1]{%
 \ifx #1\expandafter \@firstoftwo
 \else \expandafter \@secondoftwo
 \fi
}%
\providecommand \natexlab [1]{#1}%
\providecommand \enquote  [1]{``#1''}%
\providecommand \bibnamefont  [1]{#1}%
\providecommand \bibfnamefont [1]{#1}%
\providecommand \citenamefont [1]{#1}%
\providecommand \href@noop [0]{\@secondoftwo}%
\providecommand \href [0]{\begingroup \@sanitize@url \@href}%
\providecommand \@href[1]{\@@startlink{#1}\@@href}%
\providecommand \@@href[1]{\endgroup#1\@@endlink}%
\providecommand \@sanitize@url [0]{\catcode `\\12\catcode `\$12\catcode
  `\&12\catcode `\#12\catcode `\^12\catcode `\_12\catcode `\%12\relax}%
\providecommand \@@startlink[1]{}%
\providecommand \@@endlink[0]{}%
\providecommand \url  [0]{\begingroup\@sanitize@url \@url }%
\providecommand \@url [1]{\endgroup\@href {#1}{\urlprefix }}%
\providecommand \urlprefix  [0]{URL }%
\providecommand \Eprint [0]{\href }%
\providecommand \doibase [0]{http://dx.doi.org/}%
\providecommand \selectlanguage [0]{\@gobble}%
\providecommand \bibinfo  [0]{\@secondoftwo}%
\providecommand \bibfield  [0]{\@secondoftwo}%
\providecommand \translation [1]{[#1]}%
\providecommand \BibitemOpen [0]{}%
\providecommand \bibitemStop [0]{}%
\providecommand \bibitemNoStop [0]{.\EOS\space}%
\providecommand \EOS [0]{\spacefactor3000\relax}%
\providecommand \BibitemShut  [1]{\csname bibitem#1\endcsname}%
\let\auto@bib@innerbib\@empty
\bibitem [{\citenamefont {Deutsch}(1991)}]{deutsch_ETH}%
  \BibitemOpen
  \bibfield  {author} {\bibinfo {author} {\bibfnamefont {J.~M.}\ \bibnamefont
  {Deutsch}},\ }\href@noop {} {\bibfield  {journal} {\bibinfo  {journal}
  {Physical Review A}\ }\textbf {\bibinfo {volume} {43}},\ \bibinfo {pages}
  {2046} (\bibinfo {year} {1991})}\BibitemShut {NoStop}%
\bibitem [{\citenamefont {Srednicki}(1994)}]{srednicki_ETH}%
  \BibitemOpen
  \bibfield  {author} {\bibinfo {author} {\bibfnamefont {M.}~\bibnamefont
  {Srednicki}},\ }\href@noop {} {\bibfield  {journal} {\bibinfo  {journal}
  {Physical Review E}\ }\textbf {\bibinfo {volume} {50}},\ \bibinfo {pages}
  {888} (\bibinfo {year} {1994})}\BibitemShut {NoStop}%
\bibitem [{\citenamefont {Rigol}\ \emph {et~al.}(2008)\citenamefont {Rigol},
  \citenamefont {Dunjko},\ and\ \citenamefont {Olshanii}}]{rigol_ETH}%
  \BibitemOpen
  \bibfield  {author} {\bibinfo {author} {\bibfnamefont {M.}~\bibnamefont
  {Rigol}}, \bibinfo {author} {\bibfnamefont {V.}~\bibnamefont {Dunjko}}, \
  and\ \bibinfo {author} {\bibfnamefont {M.}~\bibnamefont {Olshanii}},\
  }\href@noop {} {\bibfield  {journal} {\bibinfo  {journal} {Nature}\ }\textbf
  {\bibinfo {volume} {452}},\ \bibinfo {pages} {854} (\bibinfo {year}
  {2008})}\BibitemShut {NoStop}%
\bibitem [{\citenamefont {Rigol}(2009)}]{rigol_ETH2}%
  \BibitemOpen
  \bibfield  {author} {\bibinfo {author} {\bibfnamefont {M.}~\bibnamefont
  {Rigol}},\ }\href@noop {} {\bibfield  {journal} {\bibinfo  {journal}
  {Physical Review A}\ }\textbf {\bibinfo {volume} {80}},\ \bibinfo {pages}
  {053607} (\bibinfo {year} {2009})}\BibitemShut {NoStop}%
\bibitem [{\citenamefont {Rigol}\ \emph {et~al.}(2007)\citenamefont {Rigol},
  \citenamefont {Dunjko}, \citenamefont {Yurovsky},\ and\ \citenamefont
  {Olshanii}}]{rigol_GGE}%
  \BibitemOpen
  \bibfield  {author} {\bibinfo {author} {\bibfnamefont {M.}~\bibnamefont
  {Rigol}}, \bibinfo {author} {\bibfnamefont {V.}~\bibnamefont {Dunjko}},
  \bibinfo {author} {\bibfnamefont {V.}~\bibnamefont {Yurovsky}}, \ and\
  \bibinfo {author} {\bibfnamefont {M.}~\bibnamefont {Olshanii}},\ }\href@noop
  {} {\bibfield  {journal} {\bibinfo  {journal} {Physical review letters}\
  }\textbf {\bibinfo {volume} {98}},\ \bibinfo {pages} {050405} (\bibinfo
  {year} {2007})}\BibitemShut {NoStop}%
\bibitem [{\citenamefont {Langen}\ \emph {et~al.}(2015)\citenamefont {Langen},
  \citenamefont {Erne}, \citenamefont {Geiger}, \citenamefont {Rauer},
  \citenamefont {Schweigler}, \citenamefont {Kuhnert}, \citenamefont
  {Rohringer}, \citenamefont {Mazets}, \citenamefont {Gasenzer},\ and\
  \citenamefont {Schmiedmayer}}]{schmiedmayer_GGE}%
  \BibitemOpen
  \bibfield  {author} {\bibinfo {author} {\bibfnamefont {T.}~\bibnamefont
  {Langen}}, \bibinfo {author} {\bibfnamefont {S.}~\bibnamefont {Erne}},
  \bibinfo {author} {\bibfnamefont {R.}~\bibnamefont {Geiger}}, \bibinfo
  {author} {\bibfnamefont {B.}~\bibnamefont {Rauer}}, \bibinfo {author}
  {\bibfnamefont {T.}~\bibnamefont {Schweigler}}, \bibinfo {author}
  {\bibfnamefont {M.}~\bibnamefont {Kuhnert}}, \bibinfo {author} {\bibfnamefont
  {W.}~\bibnamefont {Rohringer}}, \bibinfo {author} {\bibfnamefont {I.~E.}\
  \bibnamefont {Mazets}}, \bibinfo {author} {\bibfnamefont {T.}~\bibnamefont
  {Gasenzer}}, \ and\ \bibinfo {author} {\bibfnamefont {J.}~\bibnamefont
  {Schmiedmayer}},\ }\href@noop {} {\bibfield  {journal} {\bibinfo  {journal}
  {Science}\ }\textbf {\bibinfo {volume} {348}},\ \bibinfo {pages} {207}
  (\bibinfo {year} {2015})}\BibitemShut {NoStop}%
\bibitem [{\citenamefont {Kinoshita}\ \emph {et~al.}(2006)\citenamefont
  {Kinoshita}, \citenamefont {Wenger},\ and\ \citenamefont
  {Weiss}}]{kinoshita_newton}%
  \BibitemOpen
  \bibfield  {author} {\bibinfo {author} {\bibfnamefont {T.}~\bibnamefont
  {Kinoshita}}, \bibinfo {author} {\bibfnamefont {T.}~\bibnamefont {Wenger}}, \
  and\ \bibinfo {author} {\bibfnamefont {D.~S.}\ \bibnamefont {Weiss}},\
  }\href@noop {} {\bibfield  {journal} {\bibinfo  {journal} {Nature}\ }\textbf
  {\bibinfo {volume} {440}},\ \bibinfo {pages} {900} (\bibinfo {year}
  {2006})}\BibitemShut {NoStop}%
\bibitem [{\citenamefont {Caux}\ and\ \citenamefont
  {Essler}(2013)}]{caux_quench}%
  \BibitemOpen
  \bibfield  {author} {\bibinfo {author} {\bibfnamefont {J.-S.}\ \bibnamefont
  {Caux}}\ and\ \bibinfo {author} {\bibfnamefont {F.~H.}\ \bibnamefont
  {Essler}},\ }\href@noop {} {\bibfield  {journal} {\bibinfo  {journal}
  {Physical review letters}\ }\textbf {\bibinfo {volume} {110}},\ \bibinfo
  {pages} {257203} (\bibinfo {year} {2013})}\BibitemShut {NoStop}%
\bibitem [{\citenamefont {Berges}\ \emph {et~al.}(2004)\citenamefont {Berges},
  \citenamefont {Bors\'anyi},\ and\ \citenamefont
  {Wetterich}}]{berges_prethermalization}%
  \BibitemOpen
  \bibfield  {author} {\bibinfo {author} {\bibfnamefont {J.}~\bibnamefont
  {Berges}}, \bibinfo {author} {\bibfnamefont {S.}~\bibnamefont {Bors\'anyi}},
  \ and\ \bibinfo {author} {\bibfnamefont {C.}~\bibnamefont {Wetterich}},\
  }\href {\doibase 10.1103/PhysRevLett.93.142002} {\bibfield  {journal}
  {\bibinfo  {journal} {Phys. Rev. Lett.}\ }\textbf {\bibinfo {volume} {93}},\
  \bibinfo {pages} {142002} (\bibinfo {year} {2004})}\BibitemShut {NoStop}%
\bibitem [{\citenamefont {Bertini}\ \emph {et~al.}(2015)\citenamefont
  {Bertini}, \citenamefont {Essler}, \citenamefont {Groha},\ and\ \citenamefont
  {Robinson}}]{essler_prethermalization}%
  \BibitemOpen
  \bibfield  {author} {\bibinfo {author} {\bibfnamefont {B.}~\bibnamefont
  {Bertini}}, \bibinfo {author} {\bibfnamefont {F.~H.}\ \bibnamefont {Essler}},
  \bibinfo {author} {\bibfnamefont {S.}~\bibnamefont {Groha}}, \ and\ \bibinfo
  {author} {\bibfnamefont {N.~J.}\ \bibnamefont {Robinson}},\ }\href@noop {}
  {\bibfield  {journal} {\bibinfo  {journal} {Physical review letters}\
  }\textbf {\bibinfo {volume} {115}},\ \bibinfo {pages} {180601} (\bibinfo
  {year} {2015})}\BibitemShut {NoStop}%
\bibitem [{\citenamefont {Buchhold}\ \emph {et~al.}(2016)\citenamefont
  {Buchhold}, \citenamefont {Heyl},\ and\ \citenamefont
  {Diehl}}]{diehl_prethermalization}%
  \BibitemOpen
  \bibfield  {author} {\bibinfo {author} {\bibfnamefont {M.}~\bibnamefont
  {Buchhold}}, \bibinfo {author} {\bibfnamefont {M.}~\bibnamefont {Heyl}}, \
  and\ \bibinfo {author} {\bibfnamefont {S.}~\bibnamefont {Diehl}},\ }\href
  {\doibase 10.1103/PhysRevA.94.013601} {\bibfield  {journal} {\bibinfo
  {journal} {Phys. Rev. A}\ }\textbf {\bibinfo {volume} {94}},\ \bibinfo
  {pages} {013601} (\bibinfo {year} {2016})}\BibitemShut {NoStop}%
\bibitem [{\citenamefont {Chin}\ \emph {et~al.}(2010)\citenamefont {Chin},
  \citenamefont {Grimm}, \citenamefont {Julienne},\ and\ \citenamefont
  {Tiesinga}}]{review_feshbach}%
  \BibitemOpen
  \bibfield  {author} {\bibinfo {author} {\bibfnamefont {C.}~\bibnamefont
  {Chin}}, \bibinfo {author} {\bibfnamefont {R.}~\bibnamefont {Grimm}},
  \bibinfo {author} {\bibfnamefont {P.}~\bibnamefont {Julienne}}, \ and\
  \bibinfo {author} {\bibfnamefont {E.}~\bibnamefont {Tiesinga}},\ }\href@noop
  {} {\bibfield  {journal} {\bibinfo  {journal} {Reviews of Modern Physics}\
  }\textbf {\bibinfo {volume} {82}},\ \bibinfo {pages} {1225} (\bibinfo {year}
  {2010})}\BibitemShut {NoStop}%
\bibitem [{\citenamefont {Hung}\ \emph {et~al.}(2013)\citenamefont {Hung},
  \citenamefont {Gurarie},\ and\ \citenamefont {Chin}}]{chin_sakharov}%
  \BibitemOpen
  \bibfield  {author} {\bibinfo {author} {\bibfnamefont {C.-L.}\ \bibnamefont
  {Hung}}, \bibinfo {author} {\bibfnamefont {V.}~\bibnamefont {Gurarie}}, \
  and\ \bibinfo {author} {\bibfnamefont {C.}~\bibnamefont {Chin}},\ }\href@noop
  {} {\bibfield  {journal} {\bibinfo  {journal} {Science}\ ,\ \bibinfo {pages}
  {1237557}} (\bibinfo {year} {2013})}\BibitemShut {NoStop}%
\bibitem [{\citenamefont {Jaskula}\ \emph {et~al.}(2012)\citenamefont
  {Jaskula}, \citenamefont {Partridge}, \citenamefont {Bonneau}, \citenamefont
  {Lopes}, \citenamefont {Ruaudel}, \citenamefont {Boiron},\ and\ \citenamefont
  {Westbrook}}]{westbrook_DCE}%
  \BibitemOpen
  \bibfield  {author} {\bibinfo {author} {\bibfnamefont {J.-C.}\ \bibnamefont
  {Jaskula}}, \bibinfo {author} {\bibfnamefont {G.~B.}\ \bibnamefont
  {Partridge}}, \bibinfo {author} {\bibfnamefont {M.}~\bibnamefont {Bonneau}},
  \bibinfo {author} {\bibfnamefont {R.}~\bibnamefont {Lopes}}, \bibinfo
  {author} {\bibfnamefont {J.}~\bibnamefont {Ruaudel}}, \bibinfo {author}
  {\bibfnamefont {D.}~\bibnamefont {Boiron}}, \ and\ \bibinfo {author}
  {\bibfnamefont {C.~I.}\ \bibnamefont {Westbrook}},\ }\href {\doibase
  10.1103/PhysRevLett.109.220401} {\bibfield  {journal} {\bibinfo  {journal}
  {Phys. Rev. Lett.}\ }\textbf {\bibinfo {volume} {109}},\ \bibinfo {pages}
  {220401} (\bibinfo {year} {2012})}\BibitemShut {NoStop}%
\bibitem [{\citenamefont {Castin}\ and\ \citenamefont
  {Dum}(1998)}]{castin_dum}%
  \BibitemOpen
  \bibfield  {author} {\bibinfo {author} {\bibfnamefont {Y.}~\bibnamefont
  {Castin}}\ and\ \bibinfo {author} {\bibfnamefont {R.}~\bibnamefont {Dum}},\
  }\href@noop {} {\bibfield  {journal} {\bibinfo  {journal} {Physical Review
  A}\ }\textbf {\bibinfo {volume} {57}},\ \bibinfo {pages} {3008} (\bibinfo
  {year} {1998})}\BibitemShut {NoStop}%
\bibitem [{\citenamefont {Sinatra}\ \emph {et~al.}(2009)\citenamefont
  {Sinatra}, \citenamefont {Castin},\ and\ \citenamefont
  {Witkowska}}]{castin_coherence}%
  \BibitemOpen
  \bibfield  {author} {\bibinfo {author} {\bibfnamefont {A.}~\bibnamefont
  {Sinatra}}, \bibinfo {author} {\bibfnamefont {Y.}~\bibnamefont {Castin}}, \
  and\ \bibinfo {author} {\bibfnamefont {E.}~\bibnamefont {Witkowska}},\
  }\href@noop {} {\bibfield  {journal} {\bibinfo  {journal} {Physical Review
  A}\ }\textbf {\bibinfo {volume} {80}},\ \bibinfo {pages} {033614} (\bibinfo
  {year} {2009})}\BibitemShut {NoStop}%
\bibitem [{\citenamefont {Natu}\ and\ \citenamefont
  {Mueller}(2013)}]{natu_quench}%
  \BibitemOpen
  \bibfield  {author} {\bibinfo {author} {\bibfnamefont {S.~S.}\ \bibnamefont
  {Natu}}\ and\ \bibinfo {author} {\bibfnamefont {E.~J.}\ \bibnamefont
  {Mueller}},\ }\href@noop {} {\bibfield  {journal} {\bibinfo  {journal}
  {Physical Review A}\ }\textbf {\bibinfo {volume} {87}},\ \bibinfo {pages}
  {053607} (\bibinfo {year} {2013})}\BibitemShut {NoStop}%
\bibitem [{\citenamefont {Carusotto}\ \emph {et~al.}(2010)\citenamefont
  {Carusotto}, \citenamefont {Balbinot}, \citenamefont {Fabbri},\ and\
  \citenamefont {Recati}}]{iacopo_DCE}%
  \BibitemOpen
  \bibfield  {author} {\bibinfo {author} {\bibfnamefont {I.}~\bibnamefont
  {Carusotto}}, \bibinfo {author} {\bibfnamefont {R.}~\bibnamefont {Balbinot}},
  \bibinfo {author} {\bibfnamefont {A.}~\bibnamefont {Fabbri}}, \ and\ \bibinfo
  {author} {\bibfnamefont {A.}~\bibnamefont {Recati}},\ }\href@noop {}
  {\bibfield  {journal} {\bibinfo  {journal} {The European Physical Journal D}\
  }\textbf {\bibinfo {volume} {56}},\ \bibinfo {pages} {391} (\bibinfo {year}
  {2010})}\BibitemShut {NoStop}%
\bibitem [{\citenamefont {Menegoz}\ and\ \citenamefont
  {Silva}(2015)}]{menegoz_quench}%
  \BibitemOpen
  \bibfield  {author} {\bibinfo {author} {\bibfnamefont {G.}~\bibnamefont
  {Menegoz}}\ and\ \bibinfo {author} {\bibfnamefont {A.}~\bibnamefont
  {Silva}},\ }\href {http://stacks.iop.org/1742-5468/2015/i=5/a=P05035}
  {\bibfield  {journal} {\bibinfo  {journal} {Journal of Statistical Mechanics:
  Theory and Experiment}\ }\textbf {\bibinfo {volume} {2015}},\ \bibinfo
  {pages} {P05035} (\bibinfo {year} {2015})}\BibitemShut {NoStop}%
\bibitem [{\citenamefont {Ran{\c{c}}on}\ and\ \citenamefont
  {Levin}(2014)}]{levin_quench}%
  \BibitemOpen
  \bibfield  {author} {\bibinfo {author} {\bibfnamefont {A.}~\bibnamefont
  {Ran{\c{c}}on}}\ and\ \bibinfo {author} {\bibfnamefont {K.}~\bibnamefont
  {Levin}},\ }\href@noop {} {\bibfield  {journal} {\bibinfo  {journal}
  {Physical Review A}\ }\textbf {\bibinfo {volume} {90}},\ \bibinfo {pages}
  {021602} (\bibinfo {year} {2014})}\BibitemShut {NoStop}%
\bibitem [{\citenamefont {Beliaev}(1958)}]{BL}%
  \BibitemOpen
  \bibfield  {author} {\bibinfo {author} {\bibfnamefont {S.}~\bibnamefont
  {Beliaev}},\ }\href@noop {} {\bibfield  {journal} {\bibinfo  {journal} {Sov.
  Phys. JETP}\ }\textbf {\bibinfo {volume} {34}},\ \bibinfo {pages} {299}
  (\bibinfo {year} {1958})}\BibitemShut {NoStop}%
\bibitem [{\citenamefont {Pitaevskii}\ and\ \citenamefont
  {Stringari}(1997)}]{pitaevskii_BL}%
  \BibitemOpen
  \bibfield  {author} {\bibinfo {author} {\bibfnamefont {L.}~\bibnamefont
  {Pitaevskii}}\ and\ \bibinfo {author} {\bibfnamefont {S.}~\bibnamefont
  {Stringari}},\ }\href@noop {} {\bibfield  {journal} {\bibinfo  {journal}
  {Physics Letters A}\ }\textbf {\bibinfo {volume} {235}},\ \bibinfo {pages}
  {398} (\bibinfo {year} {1997})}\BibitemShut {NoStop}%
\bibitem [{\citenamefont {Giorgini}(1998)}]{giorgini_BL}%
  \BibitemOpen
  \bibfield  {author} {\bibinfo {author} {\bibfnamefont {S.}~\bibnamefont
  {Giorgini}},\ }\href@noop {} {\bibfield  {journal} {\bibinfo  {journal}
  {Physical Review A}\ }\textbf {\bibinfo {volume} {57}},\ \bibinfo {pages}
  {2949} (\bibinfo {year} {1998})}\BibitemShut {NoStop}%
\bibitem [{\citenamefont {Chevy}\ \emph {et~al.}(2002)\citenamefont {Chevy},
  \citenamefont {Bretin}, \citenamefont {Rosenbusch}, \citenamefont {Madison},\
  and\ \citenamefont {Dalibard}}]{dalibard_bl}%
  \BibitemOpen
  \bibfield  {author} {\bibinfo {author} {\bibfnamefont {F.}~\bibnamefont
  {Chevy}}, \bibinfo {author} {\bibfnamefont {V.}~\bibnamefont {Bretin}},
  \bibinfo {author} {\bibfnamefont {P.}~\bibnamefont {Rosenbusch}}, \bibinfo
  {author} {\bibfnamefont {K.~W.}\ \bibnamefont {Madison}}, \ and\ \bibinfo
  {author} {\bibfnamefont {J.}~\bibnamefont {Dalibard}},\ }\href {\doibase
  10.1103/PhysRevLett.88.250402} {\bibfield  {journal} {\bibinfo  {journal}
  {Phys. Rev. Lett.}\ }\textbf {\bibinfo {volume} {88}},\ \bibinfo {pages}
  {250402} (\bibinfo {year} {2002})}\BibitemShut {NoStop}%
\bibitem [{\citenamefont {Katz}\ \emph {et~al.}(2002)\citenamefont {Katz},
  \citenamefont {Steinhauer}, \citenamefont {Ozeri},\ and\ \citenamefont
  {Davidson}}]{steinhauer_damping}%
  \BibitemOpen
  \bibfield  {author} {\bibinfo {author} {\bibfnamefont {N.}~\bibnamefont
  {Katz}}, \bibinfo {author} {\bibfnamefont {J.}~\bibnamefont {Steinhauer}},
  \bibinfo {author} {\bibfnamefont {R.}~\bibnamefont {Ozeri}}, \ and\ \bibinfo
  {author} {\bibfnamefont {N.}~\bibnamefont {Davidson}},\ }\href@noop {}
  {\bibfield  {journal} {\bibinfo  {journal} {Physical review letters}\
  }\textbf {\bibinfo {volume} {89}},\ \bibinfo {pages} {220401} (\bibinfo
  {year} {2002})}\BibitemShut {NoStop}%
\bibitem [{\citenamefont {Dodonov}(2010)}]{DCE_review}%
  \BibitemOpen
  \bibfield  {author} {\bibinfo {author} {\bibfnamefont {V.}~\bibnamefont
  {Dodonov}},\ }\href@noop {} {\bibfield  {journal} {\bibinfo  {journal}
  {Physica Scripta}\ }\textbf {\bibinfo {volume} {82}},\ \bibinfo {pages}
  {038105} (\bibinfo {year} {2010})}\BibitemShut {NoStop}%
\bibitem [{\citenamefont {Koghee}\ and\ \citenamefont
  {Wouters}(2014)}]{selma_DCE}%
  \BibitemOpen
  \bibfield  {author} {\bibinfo {author} {\bibfnamefont {S.}~\bibnamefont
  {Koghee}}\ and\ \bibinfo {author} {\bibfnamefont {M.}~\bibnamefont
  {Wouters}},\ }\href@noop {} {\bibfield  {journal} {\bibinfo  {journal}
  {Physical review letters}\ }\textbf {\bibinfo {volume} {112}},\ \bibinfo
  {pages} {036406} (\bibinfo {year} {2014})}\BibitemShut {NoStop}%
\bibitem [{\citenamefont {Wilson}\ \emph {et~al.}(2011)\citenamefont {Wilson},
  \citenamefont {Johansson}, \citenamefont {Pourkabirian}, \citenamefont
  {Simoen}, \citenamefont {Johansson}, \citenamefont {Duty}, \citenamefont
  {Nori},\ and\ \citenamefont {Delsing}}]{wilson_DCE}%
  \BibitemOpen
  \bibfield  {author} {\bibinfo {author} {\bibfnamefont {C.}~\bibnamefont
  {Wilson}}, \bibinfo {author} {\bibfnamefont {G.}~\bibnamefont {Johansson}},
  \bibinfo {author} {\bibfnamefont {A.}~\bibnamefont {Pourkabirian}}, \bibinfo
  {author} {\bibfnamefont {M.}~\bibnamefont {Simoen}}, \bibinfo {author}
  {\bibfnamefont {J.}~\bibnamefont {Johansson}}, \bibinfo {author}
  {\bibfnamefont {T.}~\bibnamefont {Duty}}, \bibinfo {author} {\bibfnamefont
  {F.}~\bibnamefont {Nori}}, \ and\ \bibinfo {author} {\bibfnamefont
  {P.}~\bibnamefont {Delsing}},\ }\href@noop {} {\bibfield  {journal} {\bibinfo
   {journal} {Nature}\ }\textbf {\bibinfo {volume} {479}},\ \bibinfo {pages}
  {376} (\bibinfo {year} {2011})}\BibitemShut {NoStop}%
\bibitem [{\citenamefont {Proukakis}(2001)}]{proukakis_hierarchy}%
  \BibitemOpen
  \bibfield  {author} {\bibinfo {author} {\bibfnamefont {N.~P.}\ \bibnamefont
  {Proukakis}},\ }\href {http://stacks.iop.org/0953-4075/34/i=23/a=317}
  {\bibfield  {journal} {\bibinfo  {journal} {Journal of Physics B: Atomic,
  Molecular and Optical Physics}\ }\textbf {\bibinfo {volume} {34}},\ \bibinfo
  {pages} {4737} (\bibinfo {year} {2001})}\BibitemShut {NoStop}%
\bibitem [{\citenamefont {Van~Regemortel}\ \emph {et~al.}(2017)\citenamefont
  {Van~Regemortel}, \citenamefont {Casteels}, \citenamefont {Carusotto},\ and\
  \citenamefont {Wouters}}]{beliaev_us}%
  \BibitemOpen
  \bibfield  {author} {\bibinfo {author} {\bibfnamefont {M.}~\bibnamefont
  {Van~Regemortel}}, \bibinfo {author} {\bibfnamefont {W.}~\bibnamefont
  {Casteels}}, \bibinfo {author} {\bibfnamefont {I.}~\bibnamefont {Carusotto}},
  \ and\ \bibinfo {author} {\bibfnamefont {M.}~\bibnamefont {Wouters}},\ }\href
  {\doibase 10.1103/PhysRevA.96.053854} {\bibfield  {journal} {\bibinfo
  {journal} {Phys. Rev. A}\ }\textbf {\bibinfo {volume} {96}},\ \bibinfo
  {pages} {053854} (\bibinfo {year} {2017})}\BibitemShut {NoStop}%
\bibitem [{\citenamefont {Lipschitz}\ and\ \citenamefont
  {Pitaevskii}(1981)}]{LandauLipschitzVol10}%
  \BibitemOpen
  \bibfield  {author} {\bibinfo {author} {\bibfnamefont {E.}~\bibnamefont
  {Lipschitz}}\ and\ \bibinfo {author} {\bibfnamefont {L.}~\bibnamefont
  {Pitaevskii}},\ }in\ \href@noop {} {\emph {\bibinfo {booktitle} {Landau and
  Lifshitz Course of Theoretical Physics}}},\ Vol.~\bibinfo {volume} {10}\
  (\bibinfo  {publisher} {Pergamon Press},\ \bibinfo {address} {New York},\
  \bibinfo {year} {1981})\ Chap.\ \bibinfo {chapter} {VII}\BibitemShut
  {NoStop}%
\bibitem [{\citenamefont {Balbinot}\ \emph {et~al.}(2008)\citenamefont
  {Balbinot}, \citenamefont {Fabbri}, \citenamefont {Fagnocchi}, \citenamefont
  {Recati},\ and\ \citenamefont {Carusotto}}]{iacopo_hawking}%
  \BibitemOpen
  \bibfield  {author} {\bibinfo {author} {\bibfnamefont {R.}~\bibnamefont
  {Balbinot}}, \bibinfo {author} {\bibfnamefont {A.}~\bibnamefont {Fabbri}},
  \bibinfo {author} {\bibfnamefont {S.}~\bibnamefont {Fagnocchi}}, \bibinfo
  {author} {\bibfnamefont {A.}~\bibnamefont {Recati}}, \ and\ \bibinfo {author}
  {\bibfnamefont {I.}~\bibnamefont {Carusotto}},\ }\href@noop {} {\bibfield
  {journal} {\bibinfo  {journal} {Physical Review A}\ }\textbf {\bibinfo
  {volume} {78}},\ \bibinfo {pages} {021603} (\bibinfo {year}
  {2008})}\BibitemShut {NoStop}%
\bibitem [{\citenamefont {Carusotto}\ \emph {et~al.}(2008)\citenamefont
  {Carusotto}, \citenamefont {Fagnocchi}, \citenamefont {Recati}, \citenamefont
  {Balbinot},\ and\ \citenamefont {Fabbri}}]{iacopo_hawking2}%
  \BibitemOpen
  \bibfield  {author} {\bibinfo {author} {\bibfnamefont {I.}~\bibnamefont
  {Carusotto}}, \bibinfo {author} {\bibfnamefont {S.}~\bibnamefont
  {Fagnocchi}}, \bibinfo {author} {\bibfnamefont {A.}~\bibnamefont {Recati}},
  \bibinfo {author} {\bibfnamefont {R.}~\bibnamefont {Balbinot}}, \ and\
  \bibinfo {author} {\bibfnamefont {A.}~\bibnamefont {Fabbri}},\ }\href@noop {}
  {\bibfield  {journal} {\bibinfo  {journal} {New Journal of Physics}\ }\textbf
  {\bibinfo {volume} {10}},\ \bibinfo {pages} {103001} (\bibinfo {year}
  {2008})}\BibitemShut {NoStop}%
\bibitem [{\citenamefont {Steinhauer}(2016)}]{steinhauer_hawking}%
  \BibitemOpen
  \bibfield  {author} {\bibinfo {author} {\bibfnamefont {J.}~\bibnamefont
  {Steinhauer}},\ }\href@noop {} {\bibfield  {journal} {\bibinfo  {journal}
  {Nature Physics}\ }\textbf {\bibinfo {volume} {12}},\ \bibinfo {pages} {959}
  (\bibinfo {year} {2016})}\BibitemShut {NoStop}%
\bibitem [{\citenamefont {Calabrese}\ and\ \citenamefont
  {Cardy}(2005)}]{calabrese_qp}%
  \BibitemOpen
  \bibfield  {author} {\bibinfo {author} {\bibfnamefont {P.}~\bibnamefont
  {Calabrese}}\ and\ \bibinfo {author} {\bibfnamefont {J.}~\bibnamefont
  {Cardy}},\ }\href@noop {} {\bibfield  {journal} {\bibinfo  {journal} {Journal
  of Statistical Mechanics: Theory and Experiment}\ }\textbf {\bibinfo {volume}
  {2005}},\ \bibinfo {pages} {P04010} (\bibinfo {year} {2005})}\BibitemShut
  {NoStop}%
\bibitem [{\citenamefont {Cohen-Tannoudji}\ \emph {et~al.}(1988)\citenamefont
  {Cohen-Tannoudji}, \citenamefont {Dupont-Roc},\ and\ \citenamefont
  {Grynberg}}]{Cohen}%
  \BibitemOpen
  \bibfield  {author} {\bibinfo {author} {\bibfnamefont {C.}~\bibnamefont
  {Cohen-Tannoudji}}, \bibinfo {author} {\bibfnamefont {J.}~\bibnamefont
  {Dupont-Roc}}, \ and\ \bibinfo {author} {\bibfnamefont {G.}~\bibnamefont
  {Grynberg}},\ }\href@noop {} {\emph {\bibinfo {title} {{Processus
  d'interaction entre photons et atomes}}}}\ (\bibinfo  {publisher}
  {InterEditions et \'Editions du {CNRS}},\ \bibinfo {address} {Paris},\
  \bibinfo {year} {1988})\BibitemShut {NoStop}%
\bibitem [{\citenamefont {Kurkjian}\ \emph {et~al.}(2017)\citenamefont
  {Kurkjian}, \citenamefont {Castin},\ and\ \citenamefont {Sinatra}}]{Annalen}%
  \BibitemOpen
  \bibfield  {author} {\bibinfo {author} {\bibfnamefont {H.}~\bibnamefont
  {Kurkjian}}, \bibinfo {author} {\bibfnamefont {Y.}~\bibnamefont {Castin}}, \
  and\ \bibinfo {author} {\bibfnamefont {A.}~\bibnamefont {Sinatra}},\ }\href
  {\doibase 10.1002/andp.201600352} {\bibfield  {journal} {\bibinfo  {journal}
  {Annalen der Physik}\ }\textbf {\bibinfo {volume} {529}},\ \bibinfo {pages}
  {1600352} (\bibinfo {year} {2017})}\BibitemShut {NoStop}%
\end{thebibliography}%

\end{document}